# Deep LF-Net: Semantic Lung Segmentation from Indian Chest Radiographs Including Severely Unhealthy Images

Anushikha Singh[1], Brejesh Lall[2], B. K. Panigrahi[2], Anjali Agrawal[3], Anurag Agrawal[4], DJ Christopher[5], Balamugesh Thangakunam[5]

[1]Bharti School of Telecommunications Technology & Management, Indian Institute of Technology Delhi, New Delhi 110016, India

[2]Department of Electrical Engineering, Indian Institute of Technology Delhi, New Delhi 110016, India

[3]Teleradiology Solutions, Civil Lines, Delhi 110054, India

[4]CSIR-Institute of Genomics and Integrative Biology, New Delhi 110025, India

[5]Department of Pulmonary medicine, Christian Medical College, Vellore - 632004, India

Anushikha.Singh@dbst.iitd.ac.in, brejesh@ee.iitd.ac.in, bijayaketan.panigrahi@gmail.com, anjali.agrawal@telradsol.com, a.agrawal@igib.in, djchris@cmcvellore.ac.in,drbalamugesh@yahoo.com

*Abstract*—A chest radiograph, commonly called chest x-ray (CxR), plays a vital role in the diagnosis of various lung diseases, such as lung cancer, tuberculosis, pneumonia, and many more. Automated segmentation of the lungs is an important step to design a computer-aided diagnostic tool for examination of a CxR. Precise lung segmentation is considered extremely challenging because of variance in the shape of the lung caused by health issues, age, and gender. The proposed work investigates the use of an efficient deep convolutional neural network for accurate segmentation of lungs from CxR. We attempt an end to end DeepLabv3+ network which integrates DeepLab architecture, encoder-decoder, and dilated convolution for semantic lung segmentation with fast training and high accuracy. We experimented with the different pre-trained base networks: Resnet18 and Mobilenetv2, associated with the Deeplabv3+ model for performance analysis. The proposed approach does not require any pre-processing technique on chest x-ray images before being fed to a neural network. Morphological operations were used to remove false positives that occurred during semantic segmentation. We construct a dataset of CxR images with corresponding lung mask of the Indian population that contain healthy and unhealthy CxRs of clinically confirmed patients of tuberculosis, chronic obstructive pulmonary disease, interstitial lung disease, pleural effusion, and lung cancer. The proposed method is tested on 688 images of our Indian CxR dataset including images with severe abnormal findings to validate its robustness. We also experimented on commonly used benchmark datasets such as Japanese Society of Radiological Technology; Montgomery County, USA; and Shenzhen, China for state-of-the-art comparison. The performance of our method is tested against techniques described in the literature and achieved the highest accuracy for lung segmentation on Indian and public datasets.

This work was supported by the CRDF Global grant, OISE-17-62923, U.S. Authors express their thankfulness to DJ Christopher, and Balamugesh Thangakunam for collecting the Indian chest x-ray images. Authors also express their thankfulness to Dr. Anurag Agrawal, CSIR-IGIB, India, for coordinating the multi-institutional team, and critical inputs, and Dr. Anjali Agrawal, Teleradiology Solutions, India, for marking of lung boundary on chest x-ray images.

## I. INTRODUCTION

Of the various existent imaging modalities in health care, X-ray is the most commonly used diagnostic technique as it is widely available, low cost, non-invasive, and easy to acquire. Chest radiography is the most popular and important imaging modality used in the diagnosis of various pulmonary diseases, such as lung cancer, tuberculosis, emphysema, pneumonia, and many more [2]. Although CxR is widely used, it is one of the most challenging medical images to interpret. The precise and accurate interpretation of chest x-ray is highly dependent on the expertise of the medical professional [3-4], who remains in short supply. Therefore, a computer-aided diagnostic (CAD) tool would be useful to automate the interpretation of chest x-rays and fill in for the paucity of trained personnel. In developing and populous countries like India, CAD is a promising tool to assist overburdened medical experts in screening and diagnostic settings [5-6]. Automated and precise segmentation of anatomical structures in a chest x-ray is an important step towards developing CAD tools. Segmentation of the lung plays a vital role in CAD tools as shape, size, and area of the lung provide significant information about the manifestation of various chest diseases. Segmentation of the lung becomes challenging due to several reasons: 1. Variation in shape and size of the lung due to age, gender, and heart dimension; 2. Variance in the texture of the lung field at hilum, clavicle, ribs, and apex; 3. The manifestation of severe lung diseases or deformities in the bones, heart, and pulmonary vein/artery; 4. Presence of foreign objects like a pacemaker and other implanted devices or buttons on patient clothes [7].

In literature, a large number of methods have been proposed for the automated segmentation of lungs which are broadly divided into two categories, 1. The classical approach, and 2. Deep neural network techniques.

*1. The Classical Approach:*

According to the literature [7-8], the classical approach of lung segmentation can be divided into four categories based on the rule, pixel classification, deformable model, and hybrid techniques. Rule-based lung segmentation methods consider predefined anatomical rules and heuristic assumptions [9-13]. In the group of pixel classification algorithms, each pixel of an image is classified as a lung or a background pixel using a traditional supervised classifier trained with samples of chest x-ray with their corresponding lung mask [8, 14]. Deformable model-based techniques use joint shape and appearance sparse learning based framework to integrate scale dependent shape and appearance statistics for lung segmentation [15-20]. Some authors worked on hybrid techniques that combine rule, pixel, or active shape model for the segmentation of lung [21-22]. These classical lung segmentation techniques are easy to implement but the lung boundaries obtained may not be optimum due to the heterogeneity of lung field shapes. The accuracy of these algorithms also suffers in the setting of pulmonary diseases affecting lung texture.

*2. Deep Neural Network techniques:*

With the availability of fast speed GPUs, Deep learning techniques have been adopted for medical image analysis in the last few years with good results [23-24]. In literature, deep neural networks were applied in a wide range of tasks, such as detection of diseases, object detection, and segmentation in medical images with promising results. Recent literature uses deep learning techniques for the semantic segmentation of lung fields from chest radiographs [24-25]. In the semantic segmentation techniques, the chest radiograph is given as input to the deep neural network which classifies each pixel as lung or non-lung class. A. Kalinovsky et al. [26] proposed lung segmentation using an encoder-decoder convolutional neural network (ED-CNN) and achieved an average accuracy of 96.2% on a dataset of 354 chest radiographs. In ED-CNN, the encoder gradually reduces the spatial dimension and extracts the semantic information from the input image whereas the decoder recovers the object details and location information and produces output image with the lung probability of each pixel. R. Mohammad et al. [27] developed a deep neural network of 5 convolutional and 2 fully connected layers for the segmentation of the lung. The input image is divided into patches of different sizes which were fed to the neural networks for further classification. They reported the intersection over union (IOU) of 0.96 when tested on the Geisinger and Japanese Society of Radiological Technology (JSRT) dataset. S. Hwang et al. [28] proposed network-wise training of a deep convolutional neural network for accurate segmentation of the lung. Their algorithm is tested on the JSRT and Montgomery County (MC) datasets with a dice coefficient of 0.9800 and 0.9640 respectively. Novikov et al. [29] used the original U-Net model [30], all dropout, all convolutional, and Inverted model for multiclass segmentation of anatomical structures such as lungs, heart, clavicles on chest radiographs. They evaluated their models on JSRT data and achieved a Jaccard overlap score of 95% for lungs, 86.8% for clavicles, and 88.2% for the heart. M. Nishio et al. [31] used U-Net architecture with Bayesian optimization and achieved a Dice coefficient of 0.9790 and 0.9410 when tested on JSRT and MC dataset, respectively. They also evaluated the performance of their model on 65 additional images depicting severe abnormalities and reported a Dice coefficient of 0.8890. R. Hooda et al. [32] used versions of fully convolutional neural networks (FCN-8, FCN-16, and FCN-32) [33] for semantic segmentation of lung fields with an average accuracy of 98.92% and 97.84% on JSRT and MC datasets respectively. R. Rashid et al. [34] proposed a fully convolutional network for lung segmentation with an average accuracy of 97.1%, 97.7% & 94.2% on JSRT, MC, and a local dataset. A. Mittal et al. [35] did modification in the up-sampling process of well-known SegNet architecture [36] and trained it on the JSRT dataset. Their model achieved 98.73% accuracy when tested on JSRT and MC dataset. W. Dai. et al. [37] proposed Structure Correcting Adversarial Network for highly realistic and accurate segmentation of lungs and heart in chest radiographs. In a recent study, Chang et al. [38] applied an encoder-decoder deeplabv3+ neural network [39] with an atrous convolution for the segmentation left and right lung area separately. Their method is tested on 1736 images of NIH chest x-ray database and achieved 94.9% and 92% mean intersection over union for the two-class-model and the three-class-model, respectively. JC Souza et al. [40] used the AlexNet model for the initial segmentation of lungs using patch classification technique followed by reconstruction of lungs based on ResNet18 to refine segmentation.

As per literature, several methods have been constructed for the segmentation of lung fields and reported convincing accuracy on common benchmark datasets. To the best of our knowledge, very few works have evaluated the robustness of deep learning techniques in lung segmentation from CXR images with severe abnormal findings. All reported works on lung segmentation show good performance but highlight three important tenets for their success: Firstly, NN/learning schemes are highly dependent on datasets and hence high-quality data is critical for a successful model, secondly, higher accuracy is always appreciable in the case of medical imaging, and third, a decent performance on severely affected/critical images which makes it robust in the real world. The literature works are tested on standard datasets and that does not guarantee their performance on the Indian dataset. As our experiments reveal, they perform poorly on such samples, and therefore there is a need to develop an automated lung segmentation method for Indian chest radiographs.

The proposed work presents the efficient use of deep convolutional networks for automated semantic lung segmentation from chest radiographs. Our work contributes in the following ways:

- We applied an end-to-end Deeplabv3+ architecture integrated with dilated convolution for the segmentation of lungs with fast training and high accuracy.
- We construct the dataset of CxR images with corresponding ground truth lung mask of the Indian population including healthy and patients of various lung diseases.
- Extensive experiments on in-house Indian dataset including severe abnormal images, and common

benchmark datasets: Japanese Society of Radiological Technology (JSRT), the Montgomery County (MC), and Shenzhen datasets.

- Post-processing operations have been proposed. These techniques successfully remove falsely predicted labels that occurred during semantic segmentation.

Additionally, the proposed approach does not require any pre-processing technique to be applied on chest x-ray images before being fed to the neural network.

## II. DEEP NEURAL NETWORK FOR LUNG FIELD SEGMENTATION

Based on the available literature, deep neural networks are effectively used in the segmentation of lungs from CxR. The well-known neural networks are U-Net [30], SegNet [36], different versions of FCN [33], and Deeplab [39]. In the proposed work, we attempt the use of Deeplabv3+ architecture which is well known and superior in performance as state-of-art in the standard PASCAL VOC 2012 dataset. The DeepLabv3+ architecture integrates the encoder-decoder neural networks with spatial pyramid pooling and Atrous convolutions. Fast training and high precision are the most distinctive characteristics of Deeplabv3+ architecture. In our work, we trained Deeplabv3+ architecture associated with different base networks for segmentation of lungs from CxR hence the name "Deep LF-Net". The post-processing techniques are used to remove false positives presents in the segmented binary lung mask. We provide a brief explanation of these techniques:

### 1. Atrous Convolution:

Google's Deeplab architecture introduced the concept of Atrous convolution or also called dilated convolution to capture multi-scale information from different fields of view at the same kernel computational cost. It allows us to control the effective field of view of the convolution with the use of a parameter called Rate. It presents a generalized form of convolution as a convolution operation with rate =1 is a normal convolution. As we increase the rate, the effective field of view increases. The Deeplabv3+ architecture used in our work uses atrous convolution with rates {6, 12, and 18} to capture multi-scale information. The equation of atrous convolution for two-dimensional signals is given below:

$$y(i) = \sum_k x(i + r.k)w(k) \tag{i}$$

where $x$ input feature map, $y$ output feature map, w convolutional filter, $i$ location, and $r$ rate.

if $r = 1$ then normal convolution, and $r > 1$ for atrous convolution.

The rate $r$ decides the stride with which we sample the input data. The effective field of view of filters is adaptively controlled by the rate parameter.

### 2. Atrous Spatial Pyramid Pooling:

The Deeplabv3+ architecture employs Atrous Spatial Pyramid Pooling (ASPP) to capture multi-scale contextual information by applying several parallel atrous convolutions with multiple rates. ASPP applies 4 parallel operations: 1×1 convolution and 3×3 atrous convolution with multiple-rates {6, 12, and 18} on the input feature map, and fuse together. Fig. 1 presents the architecture of ASPP. It also uses a global average pooling to add image-level features. ASPP ensures that our model is robust enough if objects of a similar class have different scales in the image which makes it very accurate in the case of lung segmentation.

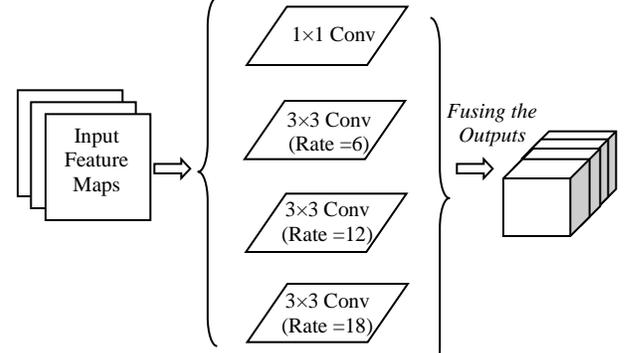

Fig. 1. Atrous Spatial Pyramid Pooling

### 3. Encoder-Decoder Neural Network:

This is a very well-known and popular architecture used in a variety of tasks in computer vision. The encoder transforms the input into a dense form that summarizes the essence of the image while the decoder recovers the spatial information. The encoder module in Deeplabv3 plus architecture employs atrous convolution to extract features at an arbitrary resolution. Here, we adopt the output stride as the ratio of the spatial resolution of an input image and final output before pooling or fully connected layer. In the proposed work of semantic lung segmentation, we consider output stride of 16 that sets the amount of encoder to downsample the input image for dense feature extraction. We experiment with two different Imagenet's pre-trained base networks, Resnet18 [41] and mobilenetv2 [42] as its main base extractor. Besides, Deeplabv3 plus uses an atrous spatial pyramid pooling module to extract convolutional features at multiple scales by employing atrous convolution with different scales. In the decoder module, the features extracted by the encoder are bi-linearly up sampled and concatenate with the low-level features after passing through the 1×1 convolution. The output features after concatenations are further convolved and bilinear sampled to refine the boundary of the semantic segmented lung field. Fig. 2 presents the architecture of Deeplabv3+ used in our work for semantic segmentation of the lung field.

### 4. Post-processing for Removal of False Positives:

After semantic segmentation of the lungs, significant false positives were present in the segmented binary lung mask. The morphological techniques have been extensively applied for noise removal in a huge variety of segmentation tasks including lung segmentation [40]. In our work, we also applied morphological operation area filtering to remove false positives. It is a well-known fact that the lungs are the two largest objects in a chest radiograph therefore area filtering is used to keep only the two largest objects which are the lung fields. All the false positives will be removed from the segmented binary lung mask for accurate segmentation. It was observed that the presence of false objects is more in the abnormal cases may be due to changes in the texture of the lungs. The post-processing technique was more useful in that case for accurate lung segmentation.

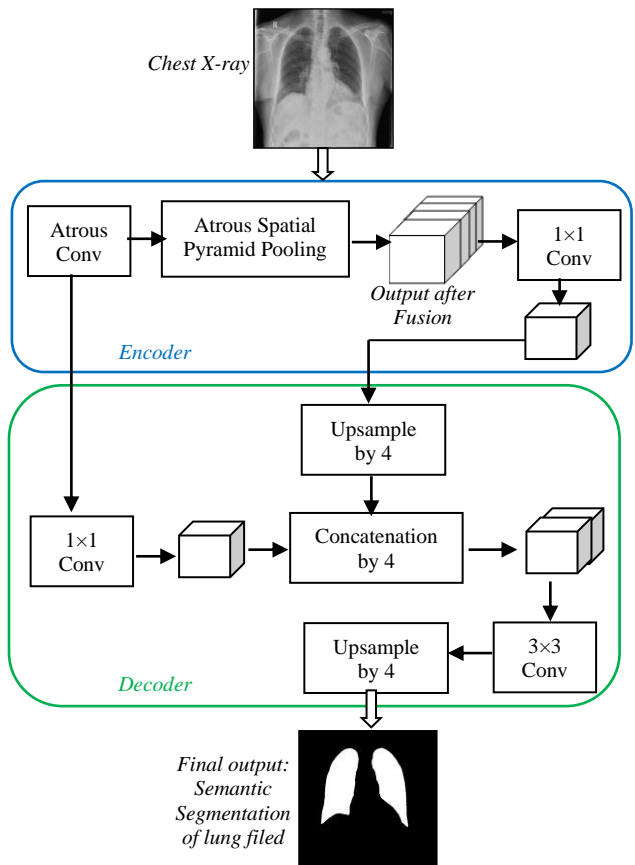

Fig. 2. The Deep LF-Net (The architecture of Deeplabv3+ used for semantic lung segmentation)

## III. Chest X-ray Datasets

The proposed work investigates the use of an efficient deep convolutional neural network for automatic segmentation of lungs from chest x-rays of the Indian population. The main challenge is the segmentation of lung regions overlapped by various dense abnormalities such as opacities, Mediastinal Adenopathy, Pleural Effusion that can be caused by different lung diseases. The presence of these dense abnormalities changes the texture of the lung region in a chest x-ray image, which can be incorrectly interpreted during the automatic lung segmentation.

The primary objective of this work is to propose a robust and accurate method of lung segmentation from the chest x-rays of the Indian population. We construct a chest x-ray dataset of the Indian population with different lung diseases and contain severe abnormal structures to validate the robustness of our method. For the generalization, we tested our method on the publically available standard datasets: Shenzhen China, Japanese Society of Radiological Technology (JSRT), and Montgomery County's, USA (MC). Table 2 presents the distribution of healthy and unhealthy chest x-ray images in the datasets used in our work. The other details of both in-house and public datasets are briefly explained below.

### A. In-House Indian Dataset:

The Indian postero-anterior (PA) CXRs were collected from Christian medical college, Vellore, India between 2014-2017.

The dataset contains a total of 688 CXRs in which 216 CXRs belong to healthy subjects and the remaining 472 CXRs are of clinically confirmed patients of different lung diseases including Tuberculosis, Chronic obstructive pulmonary disease (COPD), Interstitial lung disease (ILD), Pleural effusion (PLEF), Lung Cancer (CA). The age of the healthy subjects was 46.4±15.2 (mean ± std. deviation) and 55.6% were males. The age of the lung patients was 46.6±16.6 (mean ± std. deviation) and 69.8% were males. The TB CXRs were performed at the time of diagnosis and before or within 1 week of the institution of treatment. The digital X-rays were stored in the PACS system. All the CXRs were annotated and interpreted by experienced pulmonologists and radiologists. The boundary of lung fields is marked by our team of radiologists for validation of the proposed lung segmentation method. The format of all CXRs is JPEG and the images are of varying resolution hovering in the range of 2000 × 2000. Fig. 3 shows examples of healthy and others of lung disease chest x-ray image with the corresponding lung mask as ground truth. Table 1 presents the demographic characteristics of the subjects.

TABLE I
Demographic Characteristics of Indian Dataset

| Demographic Parameters | Healthy cases | Unhealthy cases |
|---|---|---|
| Number of patients | 216 | 472 |
| Age in years (mean ± std. deviation) | 46.4±15.2 | 46.6±16.6 |
| Gender ratio (male : female) | 55.6 : 44.4 | 69.8 : 30.2 |

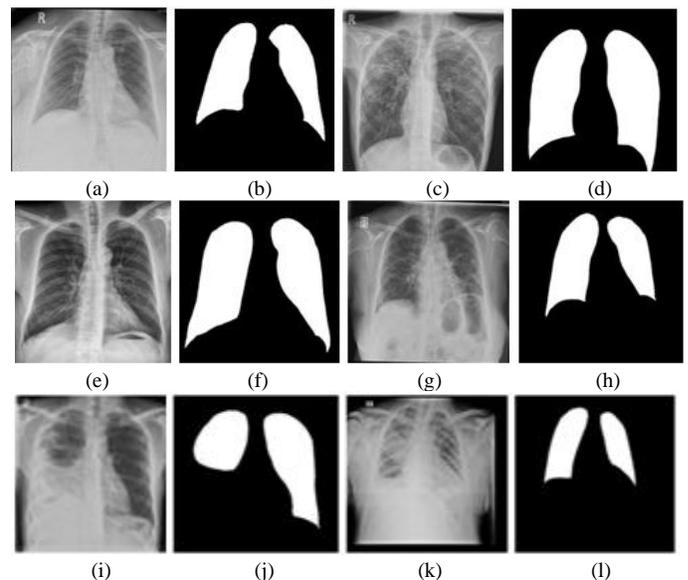

(a)  (b)  (c)  (d)

(e)  (f)  (g)  (h)

(i)  (j)  (k)  (l)

Fig. 3: Indian chest radiographs of healthy subjects and patient of different lung diseases with their corresponding lung mask.
(a) & (b) Healthy CXR with lung mask, (c) & (d) TB affected CXR with lung mask, (e) & (f) COPD affected CXR with lung mask, (g) & (h) ILD affected CXR with lung mask, (i) & (j) PLEF affected CXR with lung mask, (k) & (l) CA affected CXR with lung mask.

### B. Publically Available Standard Datasets:

In the proposed work, we used three different public datasets, JSRT, Montgomery, and Shenzhen for the state-of-the-art comparison. Below is the detailed overview of these datasets:

The JSRT dataset [43] is created by the Japanese Society of Radiological Technology in collaboration with the Japanese

Radiological Society. It contains 247 PACXRs of 2048×2048 resolution among which 154 have lung nodules and 93 have none. In our work, we used Segmented Chest Radiograph (SCR) dataset [44] as ground truth corresponds to the JSRT dataset consists of manually delineating boundaries of lungs, heart, and clavicles.

The Montgomery dataset [45] is created by the Department of Health and Human Services, Montgomery County, Maryland, USA. It contains 138 frontal CXRs including 80 healthy cases and the remaining 58 are cases of TB. The resolution of CXRs is either 4892×4020 or 4020×4892 in terms of pixels. This dataset also has lung segmentation masks as ground truth which were marked under the supervision of a radiologist and made available by Candemir et al. [46].

The images of Shenzhen dataset [45, 46] were collected by the National Library of Medicine, Maryland, USA with the collaboration of Shenzhen No.3 People's Hospital, and Guangdong Medical College, Shenzhen, China. This dataset contains 662 chest X-ray images, including 326 belonging to healthy subjects and the remaining 336 belonging to TB patients. All these images are of different resolutions approximately 3000×3000 in terms of pixels. The ground truth lung segmentation mask for the Shenzhen dataset is also publically available for performance evaluation.

TABLE II
Chest X-ray datasets used in the proposed work.

| Datasets | JSRT | MC | Shenzhen | Indian |
|---|---|---|---|---|
| Healthy cases | 93 | 80 | 326 | 216 |
| Unhealthy cases | 154: Lung nodules | 58: TB cases | 336: TB cases | 472(334:TB, 38:COPD, 39:ILD, 25:CA, 36:PLEF) |
| Total | 247 | 138 | 662 | 688 |

## IV. Training and Assessment

In the proposed work, we train the Deeplabv3+ network for the fully automated segmentation of lung fields from chest x-rays. Since the resolution of images in datasets is different therefore all the images were resized to a fixed size of 256×256 for the deep learning model. The quantitative evaluation of our method was done for each dataset discussed in the dataset section. DeepLabv3+ network is separately trained and tested for training and testing images belong to each dataset. For the evaluation, the dataset is divided into training and testing sets without overlapping with each other. 70% of total images were used for incremental training of NNs and the remaining 30% images were used for testing the model. Both training and testing datasets include healthy as well as disease affected chest x-rays for better training and generalizability. We train the Deeplabv3+ network with two different Imagenet's pre-trained base networks, Resnet18 [41] and mobilenetv2 [42] as its main feature extractor.

Deeplabv3+ model performs pixel-wise classification of lung and background class for semantic segmentation. It is obvious that pixels belonging to the lung region are less compared to the background class; therefore, we apply class weighting to balance the classes for better training.

The data augmentation techniques are performed to prevent our model from overfitting and learning the precise details of training images. The following data augmentation settings are done: 80-120% scaling, ± 10% x-axis shift, ± 10% y-axis shift, and ± 10◦ rotation.

The stochastic gradient descent is used for training with a momentum optimizer. The other hyper-parameters are a mini-batch size which is considered as 8, and the initial learning rate is selected as 0.0100. We select the downsampling factor as 16 which sets the amount to downsample the input image at the encoder section.

The proposed model is trained using the MATLAB platform on NVIDIA GeForce GTX TITAN X GPU with Intel(R) Xeon(R) CPU E5-1650 v3@3.50 GHz, 64GB RAM.

## V. Results and Discussion

This section presents the experimental results of semantic lung segmentation on in-house and public datasets achieved by the proposed method. The quantitative evaluation of our method is done using metrics extensively used and accepted for the assessment of the medical image segmentation algorithm [47]. The following metrics were used for performance evaluation of our method: Accuracy, Sensitivity, Specificity, Jaccard index or Intersection over union, and the Dice coefficient or boundary F1 score.

Results were obtained by tuning the hyper-parameters to increase the performance of the deep networks and this section reports only the most accurate results. We evaluate the performance of our model on testing data of each dataset.

This section summarizes the segmentation results of our model with two different Imagenet's pre-trained base network Resnet18 and Mobilenetv2. Table 3 comprises the results of the proposed method in terms of Accuracy, Sensitivity, Specificity, Jaccard index, and the Dice coefficient.

For the JSRT dataset, our model achieved the best accuracy of 99.37%, 99.19% of sensitivity, 99.44% of specificity, 98.50% of Jaccard Index, and 96.85% of the Dice coefficient using mobilenetv2 base network. For the MC dataset, the proposed model with the mobilenetv2 base network got the highest accuracy of 99.09%, 97.50% of sensitivity, 99.63% of specificity, 97.61% of Jaccard index, and 94.19% dice coefficient. The highest accuracy on the Shenzhen dataset is 98.31%, 97.67% sensitivity, 98.54% specificity, 95.86% Jaccard index, and 90.55% dice coefficient using mobilenetv2 base network. Deeplabv3 with a pre-trained mobilenetv2 base network provides the best performance on the public datasets.

The proposed model obtained the best accuracy of 99.52% on in-house Indian dataset with 98.95% of sensitivity, 99.71% of specificity, 95.21% of Jaccard index, and 95.28% of dice coefficient with pre-trained Resnet18 base network.

It was observed in Table 3 that there were almost no differences in the performance of our model for Indian and public datasets with both pre-trained base networks. The sensitivity is an extremely important metric to evaluate our method because it represents the rate of lung pixels correctly segmented, which is directly related to the objective of our work. Table 3 shows that the sensitivity for all public and Indian datasets is more than 97% which is very encouraging and establishes the efficiency of our method. We got more than 98% accuracy for all considered public datasets as well as on the Indian dataset which demonstrates that our model was capable of satisfactorily learning the patterns of both lung and non-lung regions in the Chest radiograph.

The random samples of segmented lung boundary for public datasets JSRT, MC, Shenzhen, and in-house Indian dataset are visualized in Fig. 3. The results illustrate that no matter how elongated, wide, or small shape of the lung is, our method is generalized enough to accurately segment the lung field in all the cases. In the Fig. 3, the ground-truth lung boundary is depicted by green, and the automatically segmented lung boundary by our method is presented by red color.

TABLE III

LUNG SEGMENTATION RESULTS ON IN-HOUSE INDIAN AND PUBLIC DATASETS.

| Metrics | Datasets | | | |
|---|---|---|---|---|
| | JSRT | MC | Shenzhen | Indian |
| *Base Network: Resnet18* | | | | |
| Accuracy | 99.31% | 99.08% | **98.31%** | **99.52%** |
| Sensitivity | 99.32% | 97.21% | 97.67% | 98.95% |
| Specificity | 99.31% | 99.72% | 98.54% | 99.71% |
| Jaccard Index | 98.38% | 97.60% | 95.86% | 95.21% |
| Dice Coefficient | 96.54% | 94.16% | 90.55% | 95.28% |
| *Base Network: Mobilenetv2* | | | | |
| Accuracy | **99.37%** | **9.09%** | 98.30% | 99.49% |
| Sensitivity | 99.19% | 97.50% | 97.63% | 98.57% |
| Specificity | 99.44% | 99.63% | 98.55% | 99.79% |
| Jaccard Index | 98.50% | 97.61% | 95.84% | 98.16% |
| Dice Coefficient | 96.85% | 94.19% | 90.54% | 97.20% |

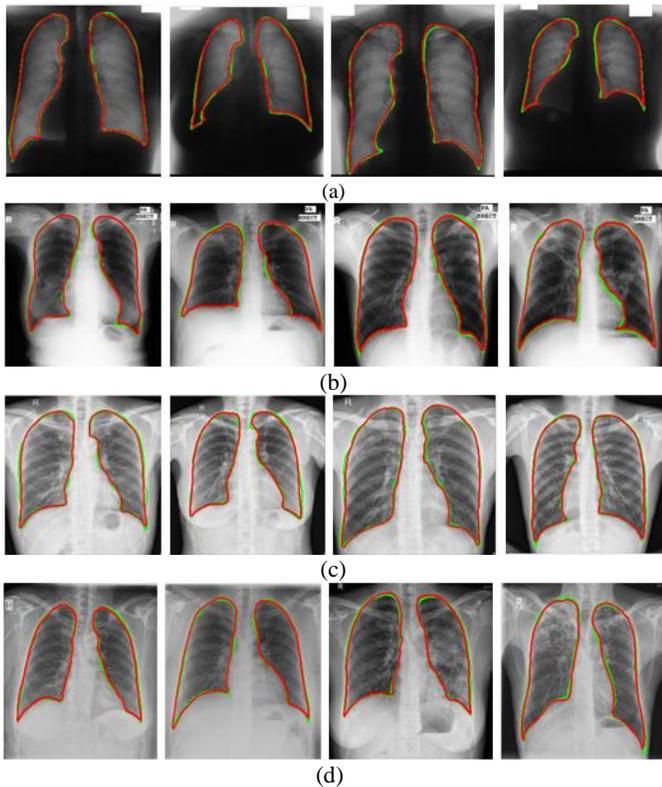

Fig. 4: Segmentation results on chest radiographs for public and Indian datasets (a) JSRT dataset, (b) MA dataset, (c) Shenzhen dataset, and (d) In-house Indian dataset.

## A. Lung Segmentation Results on Severely Unhealthy Chest Radiographs

The main challenge of lung segmentation in CXR is due to the lung region overlapped by abnormal structures in the case of severely affected CXR due to various lung diseases. Generally, these abnormal structures overlap the lung field with varying intensities, which results in a lower contrast between the lungs and their boundaries. To the best of our knowledge, very few works evaluate the robustness of their lung segmentation method on CXR images with severe abnormalities. Our in-house Indian dataset contains CXRs images of healthy subjects as well as patients of various lung diseases including TB, COPD, ILD, PLEF, and CA. We evaluate the robustness of our method on these severely affected CXRs and results are presented in this section.

Fig. 5 presents the results of lung segmentation on representative cases of CXRs with severe abnormal findings. The ground-truth lung boundary is depicted by the green, and the segmented lung boundary by our method is presented by red color. Fig. 5 shows that the detected lung boundary by our method matches very well with the ground truths. It is clear in fig. 5 that our method is robust enough to segment the lung region in the case of severe abnormal findings which makes it applicable in real-world scenarios.

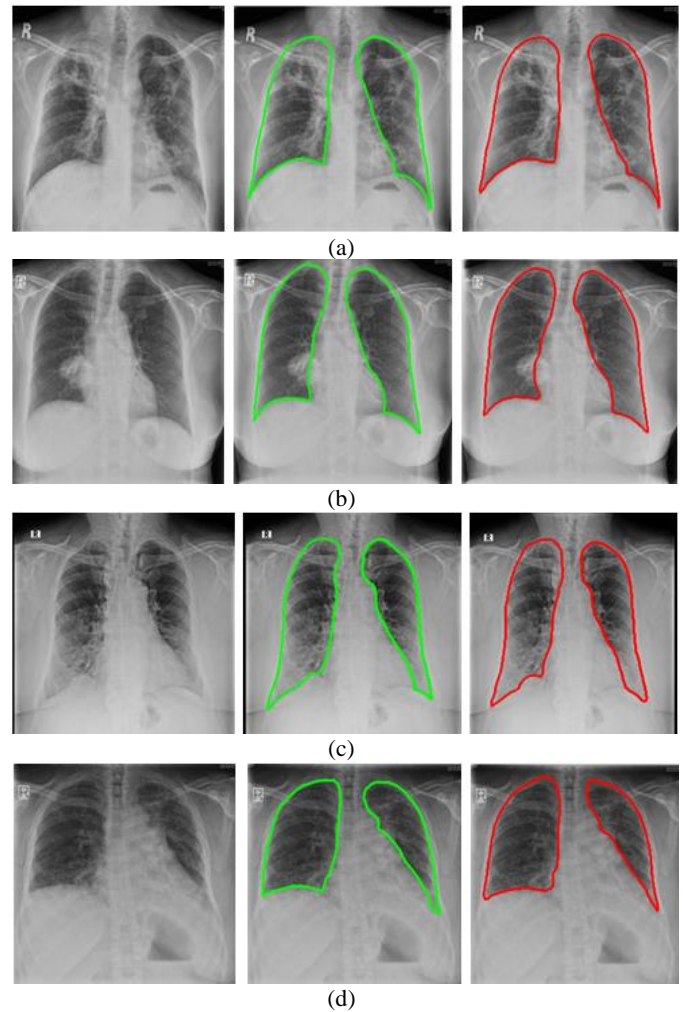

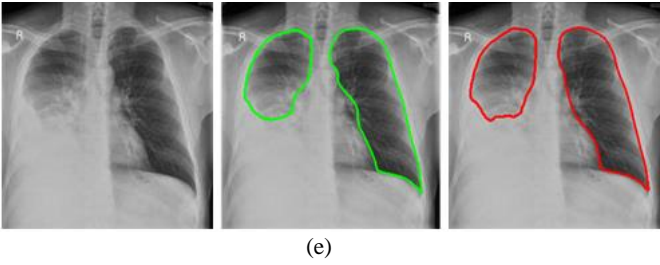

(e)

Fig. 5: Segmentation results on chest radiographs with severe abnormalities due to lung diseases (a) TB, (b) CA, (c) COPD, (d) ILD, and (e) PLEF.

### B. Failure Cases

The proposed lung segmentation method achieved more than 98% accuracy on common benchmark and in-house datasets which includes healthy and severe unhealthy images as well. Since our in-house Indian dataset contains many cases of CXR images with severe abnormal findings due to different lung diseases. We found some worst failure cases of our lung segmentation method for the Indian dataset as shown in Fig. 6. All these typical cases of CXR images belong to patients of lung diseases lung cancer, COPD, ILD, TB, and PLEF. The proposed method of lung segmentation fails due to the shadow effect of stomach air, incomplete lung region, or disappeared lung boundary because of abnormal structure. The robustness of our method on these typical cases may be improved by adding more such cases in our training data.

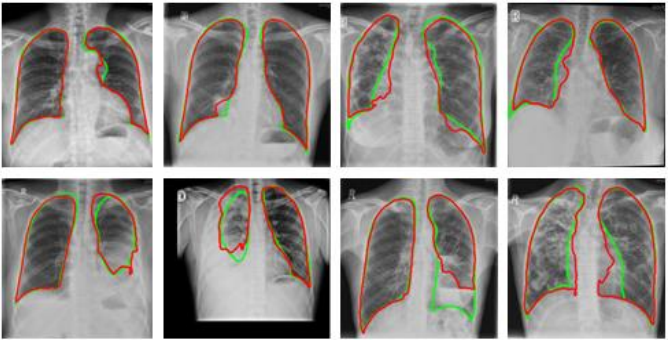

Fig. 6: Failure examples of the proposed lung segmentation method.

### C. Comparison with current state-of-the art

We tested our lung segmentation method on three common benchmark datasets for comparison with available work and validate its contribution. This state-of-the-art comparison is done with the recent literature based on deep learning techniques for the segmentation of lung field from chest x-rays. We did not consider image processing or deformable based approach of lung segmentation as deep learning surpasses these traditional techniques of segmentation [7]. Table 4 presents comparative results obtained by the proposed technique with similar literature work of lung segmentation on chest x-ray images. It can be observed in Table 4 that existing literature uses deep CNN architectures U-Net, SegNet, a different variant of FCN for pixel-wise segmentation [32, 34, 35, 49, 50], and some work is based on patch classification [27, 40] using CNN dedicated for a classification task. Table 4 compares the recent literature of lung segmentation with the method, datasets, and achieved results in all validation metrics. To the best of our knowledge, the best accuracies of semantic lung segmentation are 98.92% on the JSRT data, 97.84% on the MC data using FCN with some modification [32]. The proposed method has 99.37% accuracy with 99.32% sensitivity for lung segmentation on the JSRT dataset and 99.09% accuracy with 97.50% sensitivity on the MC dataset, outperforming the other methods. Our method provides 98.31% accuracy with 97.67% sensitivity on publicly available Shenzhen dataset, which has not been analyzed by other methods.

Table 4 shows that very few studies have evaluated the robustness of their lung segmentation method on severely abnormal images [40], a requirement in real practice. To develop a robust lung segmentation system with excellent performance on the markedly abnormal images of the in-house Indian dataset is critical, and has been addressed by our method.

Chang et al. [38] used Deeplabv3+architecture for semantic segmentation of lung on chest x-rays. They reported their results on 1736 images of the NIH chest x-ray dataset [5] and got the Jaccard Index of 94.9%. They did not evaluate the performance of their method on a common benchmark dataset therefore we could not include their work in the state of art comparison. Other deep learning based published works are included in Table 4 for comparison. It emphasizes that our proposed work reaches validation metrics superior to the existing, demonstrating the feasibility of the task of lung segmentation in a wide variety of CxRs with superlative results.

## VI. Conclusion

This paper presents an accurate and robust method for automated lung segmentation based on Deeplabv3+ architecture with different base networks mobilenetv2 and Resnet18. Our method achieved the best accuracy of 99.37%, 99.09%, and 98.31% on common benchmark datasets JSRT, MC, and Shenzhen data, respectively, outperforming the state-of-the-art methods. We constructed a CxR database of the Indian population including healthy and unhealthy images including severe lung abnormalities. The proposed method is robust and successfully segments lung fields from CxR images with varying disease severity, increasing its utility in real clinical practice.

## Acknowledgement

The study was administered under CRDF Global grant (OISE-17-62923), under the aegis of the REPORT India consortium. The consortium is supported by the Department of Biotechnology, Ministry of Science and Technology, India and the National Institute for Allergy and Infectious Diseases and National Institutes of Health (USA). DJ Christopher, and Balamugesh Thangakunam collected the Indian patients' chest X-ray images, interpreted, correlated with the clinical diagnosis & classified them. Dr. Anurag Agrawal, CSIR-IGIB, India, for coordinating the multi-institutional team, and critical inputs, and Dr. Anjali Agrawal, Teleradiology Solutions collected and labeled the images.

TABLE IV
COMPARISON OF THE PROPOSED WORK WITH RECENT LITERATURE

| Literature | Dataset | Evaluation on Severely unhealthy CXRs | Method | Performance evaluation | | | | |
|---|---|---|---|---|---|---|---|---|
| | | | | Accuracy | Specificity | Sensitivity | Dice Coefficient | Jaccard Index |
| Kalinovsky et al. [26] | JSRT + private | No | Encoder-Decoder Deep CNN | 96.2% | - | - | 97.4% | - |
| Ngo et al. [48] | JSRT | No | Deep learning + level set method | 98.5% | - | - | 99.2% | 98.5% |
| Saidy et al. [49] | JSRT | No | SegNet | - | 99.25% | 95.6% | 95.95% | |
| Hooda et al. [32] | JSRT & MC | No | Modification in FCN | 98.92% on JSRT & 97.84% on MC | - | - | - | 95.88% on JSRT & 91.74% on MC |
| CMittal et al. [35] | JSRT+ MC | | Modification in SegNet | 98.73% | | | | 95.10% |
| Rashid et al. [34] | JSRT, MC & A Local dataset of Pakistan | No | U-Net | 97.1% on JSRT, 97.7% on MC, & 94.2% Local | 98% on JSRT, 98.5% on MC, & 97% local | 95.1% on JSRT, 95.4% on MC, & 86.2% local | 95.1% on JSRT, 95.4% on MC, & 88% local | - |
| Souza et al. [40] | MC | Yes | AlexNet for patch classification + reconstruction based on ResNet18 | 96.97% | 96.79% | 97.54% | 94% | 88.07% |
| Arbabshirani et al. [27] | Geisinger & JSRT | No | CNN based patch classification | 98% on Geisinger | 99% | 99% | 96% | 91% |
| Hwang et al. [28] | JSRT & MC | No | Network-wise training of CNN | - | - | - | 98% on JSRT & 96.4% on MC | 96.1% on JSRT & 94.1% on MC |
| Novikov et al. [50] | JSRT | | FCN & Inverted-Net | - | - | - | 97.4% | 94.9% |
| The proposed work | JSRT | Yes | Deeplabv3+ architecture with pre-trained base network Resnet18 & Mobilenetv2 | 99.37% | 99.44% | 99.32% | 96.85% | 98.50% |
| | MC | | | 99.09% | 99.72% | 97.50% | 94.19% | 97.61% |
| | Shenzhen | | | 98.31% | 98.55% | 97.67% | 90.55% | 95.86% |
| | In-house Indian dataset | | | 99.52% | 99.79% | 98.95% | 97.20% | 98.16% |